\documentclass[preprint2]{aastex}
\usepackage{graphicx}
\usepackage{epstopdf}

\shorttitle{Shedding light on HLX-1}
\shortauthors{Mapelli et al.}

\begin{document}


\title{Shedding light on the nature of ESO 243-49 HLX-1}



\author{Michela Mapelli,\altaffilmark{1}}
\affil{\altaffilmark{1}INAF-Osservatorio Astronomico di Padova, Vicolo dell'Osservatorio 5, I--35122, Padova, Italy} 

\author{Francesca Annibali,\altaffilmark{2}}
\affil{\altaffilmark{2}INAF-Osservatorio Astronomico di Bologna, Via Ranzani 1, I--40127 Bologna, Italy}

\and

\author{Luca Zampieri\altaffilmark{1}}





\begin{abstract}
The point-like X-ray source HLX-1, close to the S0 galaxy ESO 243-49, is one of the strongest intermediate-mass black hole candidates. We discuss the hypothesis that ESO 243-49 is undergoing a minor merger with a gas-rich disc galaxy.  We propose that the counterpart of HLX-1 coincides with the nucleus of the secondary galaxy. We re-analyze the available photometric {\it HST} data, and we compare them with the results of N-body/smoothed particle hydrodynamics simulations. 
In particular, we derive synthetic surface brightness profiles for the simulated counterpart of HLX-1 in six {\it HST} filters, ranging from far ultraviolet (FUV) to infrared wavelengths. Such synthetic profiles include a contribution from the stellar population associated with the simulated disrupted satellite and a contribution from an irradiated disc model. These are in agreement with the observed   surface brightness profiles of the HLX-1 counterpart, provided that  the merger is at sufficiently late stage ($\gtrsim{}2.5$ Gyr since the first pericentre passage). The main difference between models and observations is in the FUV band, where the {\it HST} image shows a fuzzy and extended emission.
\end{abstract}





\section{Introduction}
\begin{figure}
\epsscale{1.1}
\plotone{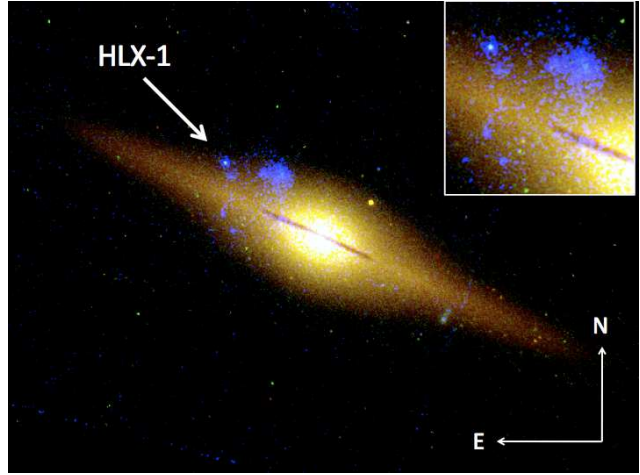}
\caption{Color-composite image of ESO~243-49. Red is F775W, green is F390W, blue is 
F140LP. The arrow indicates the HLX-1 counterpart. North is up and east is left. The insert  is a zoom ($8\times{}8$ arcsec) around the HLX-1 counterpart and the blue background galaxy at $z=0.03$. A Gaussian smoothing with $\sigma \sim 0.1$ arcsec  has been applied to the FUV image.\label{fig1}}
\end{figure}
The point-like X-ray source HLX-1, with a maximum luminosity $\sim{}10^{42}$ erg s$^{-1}$ (Farrell et al. 2009), is the brightest known ultraluminous X-ray source (ULX) and one of the strongest intermediate-mass black hole (IMBH) candidates. 
HLX-1 is located in the outskirts of the S0 galaxy ESO~243-49 (luminosity distance $\sim{}96$ Mpc), $\sim{}0.8$ kpc out of the plane and $\sim{}3.3$ kpc away from the nucleus.  The nature of the HLX-1 counterpart is still debated. Fits to the {\it Hubble Space Telescope} ({\it HST}) data (Farrell et al. 2012) indicate a total stellar mass of $\le{}6\times{}10^6$ M$_\odot{}$. Both a young ($\sim{}10$ Myr) and a old ($\sim{}10$ Gyr) age are possible.

It is not easy to explain how  a  $\sim{}10^4$ M$_\odot{}$ IMBH can form in a $\sim{}10^6$ M$_\odot{}$ star cluster (SC). Repeated mergers between stellar black holes and other compact objects and/or stars were proposed to produce IMBHs in globular clusters (e.g. Miller $\&$ Hamilton 2002), while runaway collapse of massive stars can occur in young massive SCs (e.g., Portegies Zwart \&{} McMillan 2002). An alternative scenario predicts that the IMBH is associated with the nucleus of a satellite galaxy which is being disrupted in a minor merger with  ESO~243-49 (Soria et al. 2010, 2012; Webb et al. 2010; Mapelli et al. 2012, hereafter M12). In this case, the IMBH would belong to the low-mass tail of the distribution of super massive black holes (SMBHs), located at the centre of galaxies. There is increasing evidence of  galaxy nuclei hosting  SMBHs with mass $\lesssim{}10^5$ M$_\odot{}$ (e.g. Filippenko \&{} Sargent 1989; 
Barth et al. 2004; Greene \&{} Ho 2004
Secrest et al. 2012, and references therein). The hypothesis that  ESO~243-49 recently underwent a minor merger is consistent with various features, such as the presence of prominent dust lanes around its nucleus  (e.g., Shabala et al. 2012) and the evidence of UV emission centred on its bulge, suggesting ongoing star formation (Kaviraj et al. 2009).

In the current paper, we investigate the minor-merger scenario, by analyzing a  high-resolution $N-$body/smoothed particle hydrodynamics (SPH) simulation, and by comparing the simulation outputs with the {\it HST} data. In particular, we compare the surface brightness profiles (SBPs) estimated from the simulation with the observations, from infrared to ultraviolet (UV) bands. 
\section{HST data revisited}
We re-analyzed the {\it HST} data presented by F12 (GO program 12256). The data were acquired with the Advanced Camera for Surveys (ACS) Solar Blind Camera (SBC) in the F140LP filter (far UV, FUV), and with the Wide Field Camera 3 (WFC3) in the F300X (near UV, NUV), F390W (C), F555W (V), F775W (I), and F160W (H) filters. 
The FUV, C, I color-combined image of ESO~243-49 is shown in Fig.~\ref{fig1}. It shows a fuzzy emission in FUV associated with the HLX-1 counterpart. Two `tails' departing from the source and leading south and west are visible in the image. One of the tails seems to be connected to a galaxy which was identified with a background galaxy at $z\sim{}0.03$ by Wiersema et al. (2010). This fact may lead to the conclusion that there is some connection between the counterpart of HLX-1 and the background galaxy. However, such a direct connection seems unlikely because the only line visible in the optical spectrum of the counterpart (interpreted as H$\alpha{}$) indicates a redshift for the HLX-1 counterpart closer to that of ES0~243-49 ($z=0.0224$) than to that of the background galaxy ($z\sim{}0.03$, see also Soria \&{} Hau 2012). 
\begin{figure}
\epsscale{1.1}
\plotone{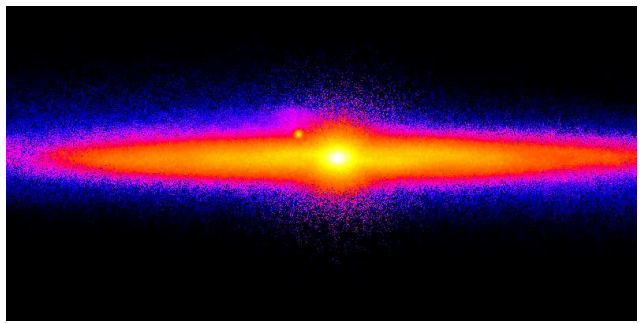}
\caption{Projected mass density of stars in the simulation at $t=2.0$ Gyr after the first pericentre passage. The primary galaxy is seen edge-on. The scale of the color-coded map is logarithmic, ranging from  $2.2\times{}10^{-6}$ M$_\odot{}$ pc$^{-3}$ to $22$ M$_\odot{}$ pc$^{-3}$.  The frame measures $50\times{}25$ kpc. \label{fig2}}
\end{figure}
\begin{figure*}
\epsscale{0.65}
\plotone{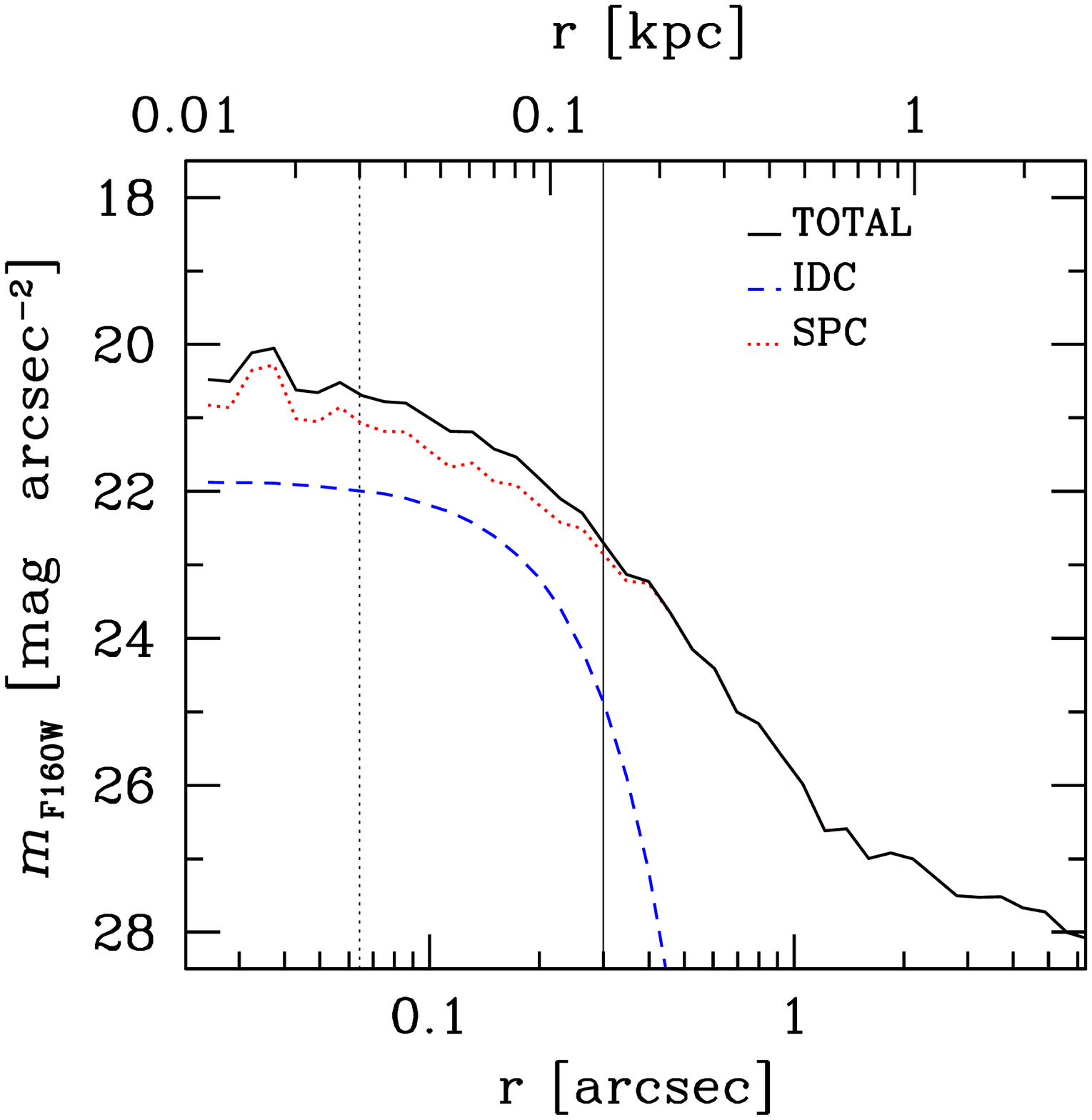}
\plotone{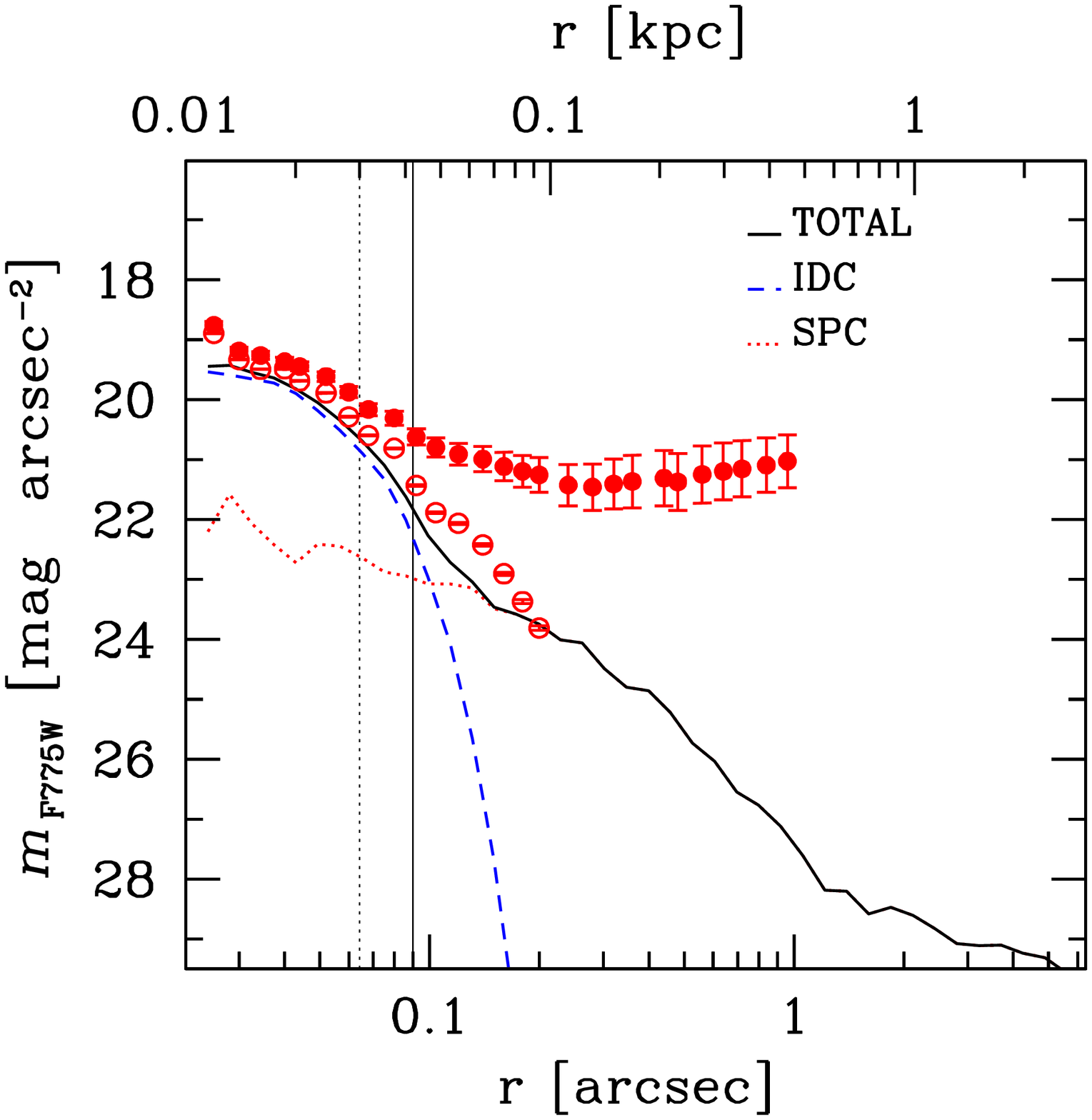}
\plotone{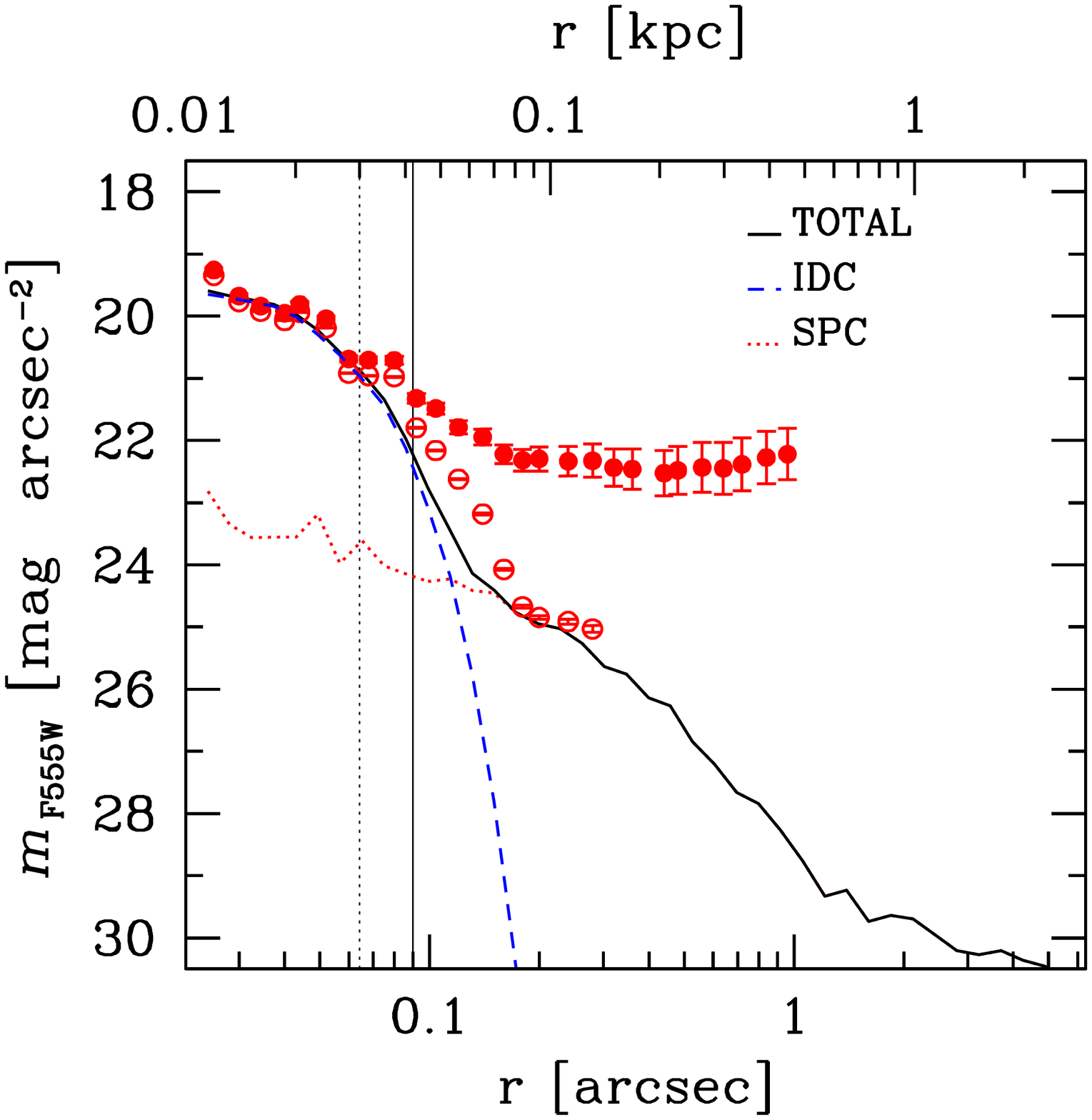}
\caption{Surface brightness profiles (SBPs) of the HLX-1 counterpart. From left to right: F160W, F775W and F555W. Circles: {\it HST} data (filled circles: first approach; open circles: second approach). Dotted red line:  stellar population component (SPC), that is simulated SBP at $t=2.5$ Gyr after the first pericentre passage; dashed blue line: irradiated disc component (IDC), smoothed over a two-dimensional Gaussian profile with the same full width at half maximum (FWHM) as the point spread function (PSF); solid black line: total SBP from the model. Vertical dotted line: softening length. Vertical solid line: PSF FWHM.\label{fig3}}
\end{figure*}

The SBPs in 5 different filters, from F140LP to F775W, were derived performing photometry within circular apertures of increasing radii (up to 1 arcsec), and computing the background in an annulus of  width 0.08 arcsec at $r>1$ arcsec (in the following, we refer to this procedure as first approach). 
In F555W and F775W, because of the high S0 flux, we attempted also a different approach, and tried to compute the SBPs after subtraction of the galaxy emission (via Gaussian smoothing with $\sigma=0.4$ arcsec). 
In the following, we refer to this procedure as second approach. We do not attempt to extract a SBP from the F160W filter, as the contamination by the S0 is too high.

\section{Model and simulations}
\begin{figure*}
\epsscale{0.65}
\plotone{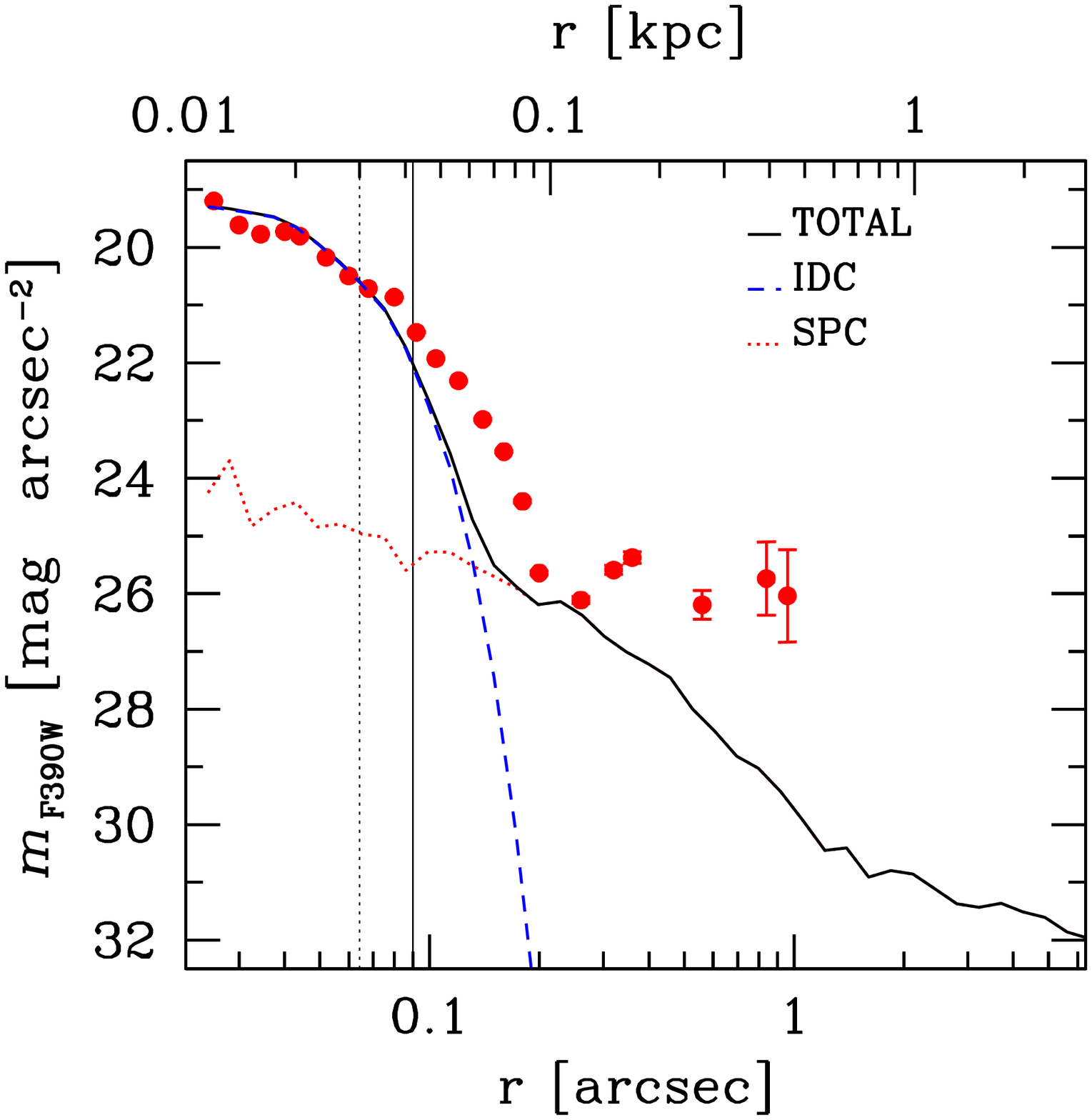}
\plotone{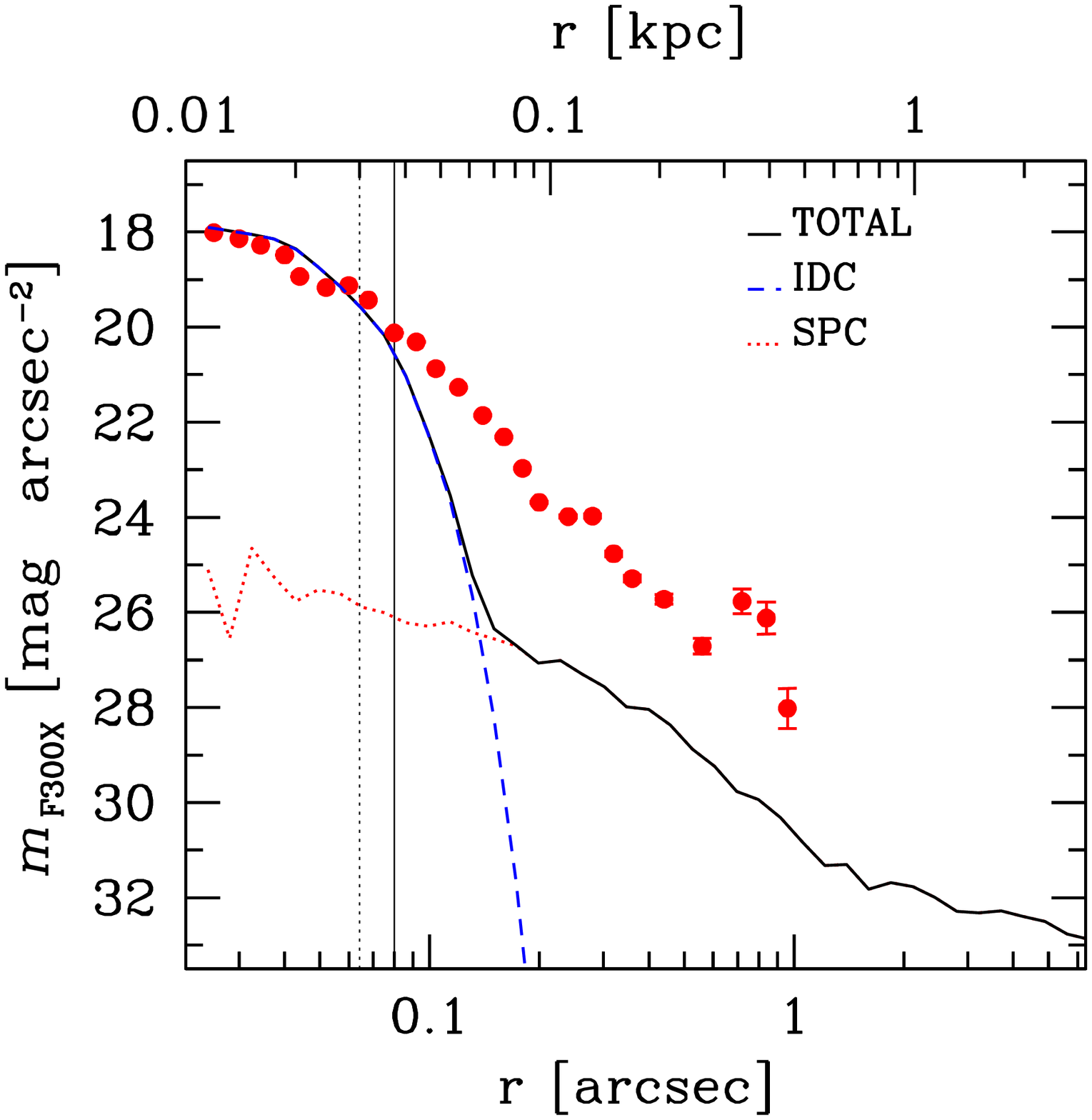}
\plotone{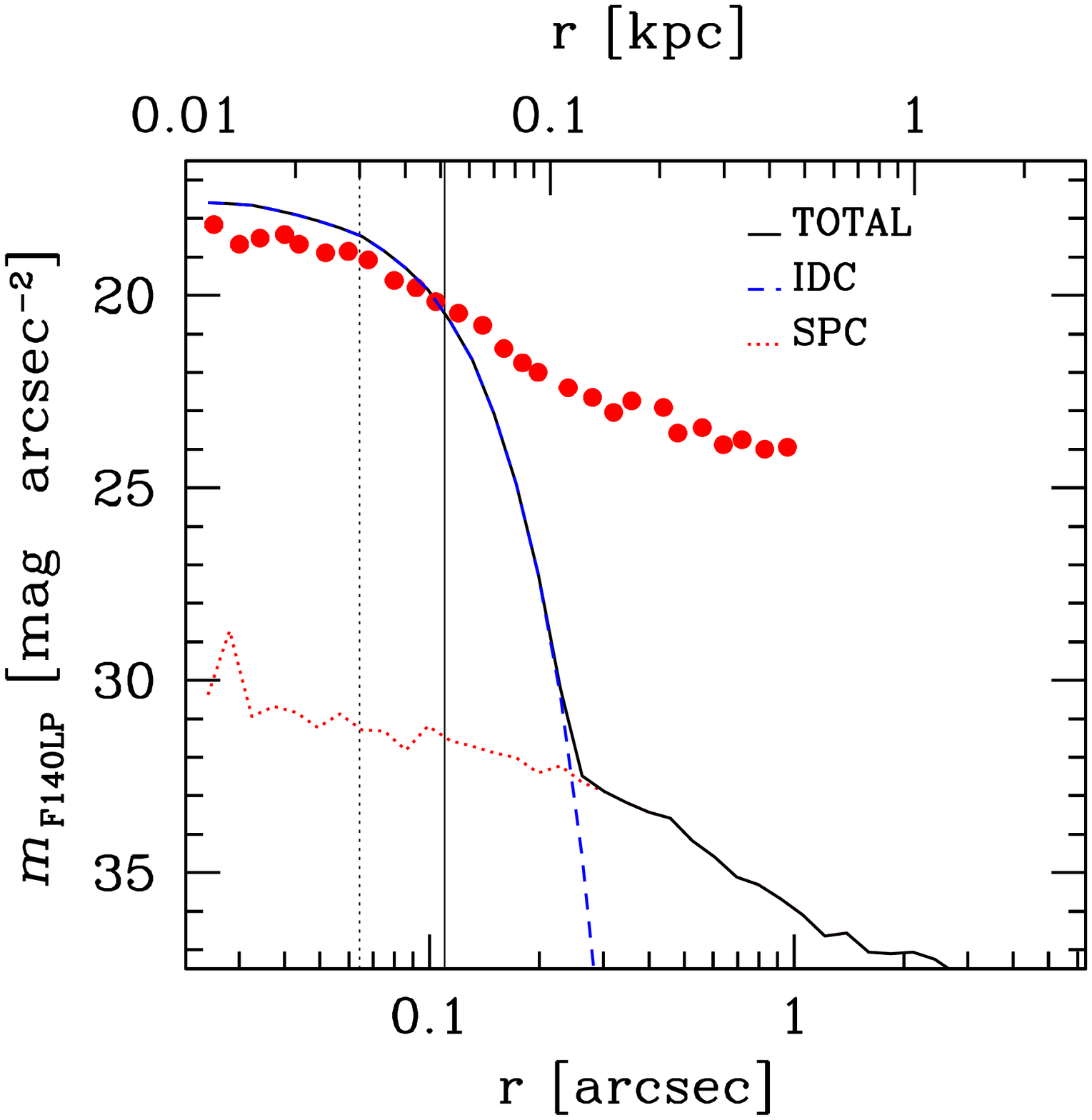}
\caption{SBPs of the HLX-1 counterpart. From left to right: F390W, F300X and F140LP. Filled circles: {\it HST} data (first approach). The lines are the same as in Fig.~\ref{fig3}, but for the three bluer filters.\label{fig4}}
\end{figure*}
{\bf $N-$body simulation:} As in M12, the initial conditions for both the primary galaxy and the secondary galaxy in the $N-$body model are generated by using an upgraded version of the code described in Widrow, Pym \&{} Dubinski (2008). 
Both the primary and the secondary galaxy have a dark matter (DM) halo, a stellar bulge and a stellar disc. The giant S0 galaxy has no gas, whereas the secondary galaxy has an initial gas mass of $1.38\times{}10^{8}$ M$_\odot{}$, distributed according to an exponential disc. The total mass of the secondary is $\sim{}1/20$ of the mass of the primary, classifying the outcome of the interaction as a minor merger. The orbit of the satellite is nearly parabolic, with initial relative speed $50$ km s$^{-1}$. 
As in M12, we simulate the evolution of the models with the $N-$body/SPH tree code gasoline (Wadsley, Stadel \&{} Quinn 2004). 
In the simulation, the particle mass in the primary galaxy is $2.5\times{}10^4$ M$_\odot{}$ and $5\times{}10^3$ M$_\odot{}$ for DM and stars, respectively. The particle mass in the secondary galaxy is  $2.5\times{}10^3$ M$_\odot{}$ for DM and $5\times{}10^2$ M$_\odot{}$ for both stars and gas. 
The softening length is $\epsilon{}=30$ pc. 
Fig.~\ref{fig2} 
shows the projected density of stars in the simulation  at $t=2.0$ Gyr after the first pericentre passage.
We can derive synthetic fluxes  from the $N-$body simulations, on the basis of each particle mass and age. In particular, we use the single stellar population (SSP) models based on the tracks of Marigo et al. (2008)
\footnote{\tt http://stev.oapd.inaf.it/cgi-bin/cmd\_2.3}. The tables of the SSP integrated magnitudes were implemented in the TIPSY visualization package for $N-$body simulations\footnote{\tt http://www-hpcc.astro.washington.edu/tools/tipsy/tipsy.html}.

{\bf Irradiated disc component (IDC):} We add an IDC to the synthetic profiles obtained from the $N-$body simulation. The IDC was modelled with a code developed for computing the optical luminosity of ULX binaries (see Patruno \&{} Zampieri  2008, for more details about the code). A standard Shakura-Sunyaev disc is assumed and both the X-ray irradiation of the companion and the self-irradiation of the disc are accounted for. 
The model shown in Figs.~\ref{fig3} and \ref{fig4} assumes IMBH mass $10^4$ M$_\odot$, bolometric luminosity $L=10^{42}$ erg s$^{-1}$, inner radius $r_{\rm in}=3\,{}r_{\rm g}$ (where $r_{\rm g}$ is the gravitational radius), outer radius  $r_{\rm out}=3.5\times{}10^{13}$ cm, inclination of $45^{\circ{}}$ and albedo 0.7. 
\section{Discussion and conclusions}
Fig.~\ref{fig3} shows that there is good agreement between the observed magnitude of the HLX-1 counterpart and the model that combines SPC and IDC,  in the filters F775W and F555W.  In the F160W filter, the total magnitude of the model (22.87) is lower than the observed one (23.49$\pm{}0.26$). 
The matching in the F160W filter can be slightly improved (22.91 versus 23.49$\pm{}0.26$) by accounting for a realistic reddening ($A_{\rm V}=0.18$). In the bluer filters (especially F300X and F140LP) the observed SBP shows a tail at $r\gtrsim{}0.2$ arcsec that extends beyond the PSF limits, cannot be due to the S0 galaxy and cannot be explained with the young stellar population in the simulated satellite (Fig.~\ref{fig4}). 

In summary, the comparison between the {\it HST} photometric data of ESO~243-49 and the results of our simulations confirms that a minor merger is a viable scenario to explain the properties of HLX-1 and of its counterpart. 
 The nature of the extended FUV and NUV emission deserves further investigation, as well as the existence of tidal tails surrounding the counterpart of HLX-1.

\acknowledgments{\scriptsize{We thank the authors of gasoline (especially J. Wadsley, T. Quinn and J. Stadel), L.~Widrow for providing us the code to generate the initial conditions, L.~Girardi and A.~Bressan for providing the SSP models. We also thank E. Ripamonti, L. R. Bedin and L. Mayer for useful discussions. We acknowledge the CINECA Award N. HP10CLI3BX and HP10B3BJEW, 2011. MM and LZ acknowledge financial support from INAF through grant PRIN-2011-1. LZ acknowledges financial support from ASI/INAF grant no. I/009/10/0.}}






\end{document}